\RequirePackage{fix-cm}
\documentclass[smallcondensed]{svjour3}     
\smartqed  
\usepackage{graphicx}
\usepackage{bm}
\usepackage{epsfig}
\usepackage{psfrag}
 \usepackage{mathptmx}      
\usepackage{amsmath}
\usepackage{amssymb}
\usepackage{latexsym}
\begin{document}

\title{$\overline{D^0} D^0$-production in $p\bar{p}$-collisions within a double handbag approach
}


\author{A.~T.~Goritschnig \and B.~Pire \and W.~Schweiger 
}


\institute{A.~T.~Goritschnig \and B.~Pire \at
              Centre de Physique Th\' eorique, \' Ecole Polytechnique \\
              91128 Palaiseau cedex, France \\
              \email{alexander.goritschnig@uni-graz.at}           
           \and
           W.~Schweiger \at
              Institute of Physics - Theory Division, University of Graz \\
              Universit\"atsplatz 5, 8010 Graz, Austria
}

\date{Received: date / Accepted: date}

\maketitle

\begin{abstract}
We estimate the scattering amplitude of the process $ p \bar{p}
\,\to\, \overline{D^0} D^0$ within a  double-handbag framework where
transition distribution amplitudes, calculated through an overlap
representation, factorize from a hard subprocess. This process will
be measured in the $\overline{\text{P}}\text{ANDA}$ experiment at GSI-FAIR.
\end{abstract}
\section{Introduction}
\label{sec:Introduction}

The $\overline{\text{P}}\text{ANDA}$ detector \cite{Lutz:2009ff} at
the Facility for Antiproton and  Ion Research (FAIR) in Darmstadt,
Germany, will provide the ideal experimental setup to study several
exclusive final channels in proton-antiproton collisions. Also the
measurement of the production of heavy hadron pairs emerging from
$p\bar{p}$-annihilations is planned, for which there is thus the
need to have theoretical input. In Ref.~\cite{Goritschnig:2012vs} we
have studied the meson-pair production $p \, \bar{p} \,\to\,
\overline{D^0} \, D^0$ within a double handbag approach in the same
line as the study of the process
 $p \, \bar{p} \,\to\, \Lambda_c^+ \, \overline{\Lambda_c}^-$ in
 Ref.~\cite{Goritschnig:2009sq},
where arguments have been given in favor of a generalization of the
handbag approach, the validity of which has been demonstrated in the
case of deeply virtual  Compton scattering and meson production. We
argue that, having the heavy $c$-quark mass $m_c$ as large intrinsic
scale and assuming restricted parton virtualities and intrinsic
transverse momenta, the $p \, \bar{p} \,\to\, \overline{D^0} \, D^0$
amplitude factorizes into a hard subprocess amplitude and soft
hadronic transition matrix elements. Neglecting intrinsic proton
charm contributions, the production of the $\bar{c}c$-pair can only
occur in the partonic subprocess. The hadronic transition matrix
elements are transition distribution amplitudes \cite{PSS}, which
generalize the concept of generalized parton distributions for
$3$-quark operators. We have developed an overlap representation
according to Ref.~\cite{Diehl:2000xz} to model the hadronic
transition matrix elements in terms of hadronic light-cone wave
functions. In our studies we only consider the valence Fock-state
components of the hadrons and we describe the proton  as  a
quark-diquark system, where we only take the scalar diquark into
account. Having the $p \, \bar{p} \,\to\, \overline{D^0} \, D^0$
amplitude at hand we are able to predict differential and integrated
$p \, \bar{p} \,\to\, \overline{D^0} \, D^0$ cross sections.

\section{Double handbag mechanism}
\label{sec:Double-handbag-mechanism}

\begin{figure}
\begin{center}
  \includegraphics[width=0.65\textwidth]{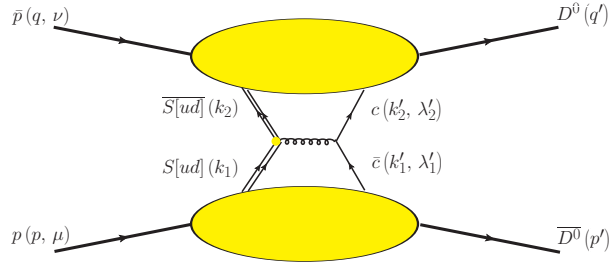}
\end{center}
\caption{The double handbag mechanism for the $p \bar{p} \,\to\,
\overline{D^0} D^0$ scattering amplitude. Blobs represent transition
distribution amplitudes (TDAs).} \label{fig:ppbar-to-DDbar}
\end{figure}
The assignment of particle momenta and helicities can be seen in
Fig.~\ref{fig:ppbar-to-DDbar}. We work in a symmetric
center-of-momentum system (CMS) in which the 3-axis is chosen to be
along the 3-vector component $\mathbf{\bar{p}}$ of the average
momentum $\bar{p} := (1/2) ( p + p^\prime )$. The transverse
momentum transfer 
 ${\bm \Delta}_\perp :=
(\Delta^1,\,\Delta^2)$, where $\Delta := p^\prime - p$, is
symmetrically shared between the incoming and the outgoing hadron
momenta. In light-front coordinates we can write the particle
momenta as follows:
\begin{equation}
\begin{split}
p \,=\,&
\left[ (1+\xi)\bar{p}^+,\, \frac{m^2+\boldsymbol{\Delta}_\perp^2/4}{2(1+\xi)\bar{p}^+},\, -\frac{{\bm \Delta}_\perp}{2} \right] \,,\quad
p^\prime \,=\,
\left[ (1-\xi)\bar{p}^+,\, \frac{M^2+\boldsymbol{\Delta}_\perp^2/4}{2(1-\xi)\bar{p}^+},\, +\frac{{\bm \Delta}_\perp}{2} \right] \,,\\
q \,=\,&
\left[ \frac{m^2+\boldsymbol{\Delta}_\perp^2/4}{2(1+\xi)\bar{p}^+},\, (1+\xi)\bar{p}^+,\, +\frac{{\bm \Delta}_\perp}{2} \right] \,,\quad
q^\prime \,=\,
\left[ \frac{m^2+\boldsymbol{\Delta}_\perp^2/4}{2(1-\xi)\bar{p}^+},\, (1-\xi)\bar{p}^+,\, -\frac{{\bm \Delta}_\perp}{2} \right] \,,
\label{eq:particle-momenta}
\end{split}
\end{equation}
where we have introduced the  skewness parameter $\xi \, := \, (-
\Delta^+)/(2\bar{p}^+)$ which parameterizes the relative momentum
transfer into longitudinal light-cone plus direction.

We consider the process $p \bar{p} \,\to\, \overline{D^0} D^0$
within a perturbative QCD  motivated framework where the hadronic
amplitude can be split up into a hard partonic subprocess and into
soft hadronic matrix elements of partonic field operators.
Specifically, we investigate the process $p \bar{p} \,\to\,
\overline{D^0} D^0$ in a  double handbag mechanism as shown in
Fig.~\ref{fig:ppbar-to-DDbar}. In such a framework only the minimal
number of hadronic constituents which are required to convert the
initial $p\bar{p}$ into the final $\overline{D^0}D^0$ pair actively
take part in the partonic subprocess. Working in a quark-scalar
diquark picture for the proton, we consider the partonic subprocess
$S[ud]\overline{S[ud]} \,\to\, \bar{c}c$. The remaining partons
inside the parent hadrons  act as spectators. In order to produce
the heavy $\bar{c}c$-pair in the partonic subprocess the gluon has
to be a highly virtual one. The $c$-quark mass serves as a natural
hard scale, allowing us to treat $S[ud]\overline{S[ud]} \,\to\,
\bar{c}c$ perturbatively . Thus, the hadronic $p \bar{p} \,\to\,
\overline{D^0} D^0$ amplitude can  be written as
\begin{equation}
\begin{split}
 M_{\mu\nu} \,=\, &
 \int d^4\bar{k}_1 \theta(\bar{k}_1^+) \int \frac{d^4 z_1}{(2\pi)^4} e^{\imath \bar{k}_1 \cdot z_1}
 \int d^4\bar{k}_2 \theta(\bar{k}_2^-) \int \frac{d^4 z_2}{(2\pi)^4} e^{\imath \bar{k}_2 \cdot z_2} \\
&\times \langle \overline{D^0} :\, p^\prime \mid \mathcal{T} \Psi^c(-z_1/2) \Phi^{S[ud]}(+z_1/2) \mid p :\, p,\,\mu \rangle
 \,\Tilde{H}(\bar{k}_1,\, \bar{k}_2)\, \\
&\times \langle D^0 :\, q^\prime \mid \mathcal{T}
\Phi^{S[ud]\,\dagger}(+z_2/2) \overline{\Psi}^c(-z_2/2) \mid \bar{p}
:\, q,\,\nu \rangle \, , \label{eq:a-priori-DHM-amplitude}
\end{split}
\end{equation}
where we have omitted color and spinor labels for the ease of
writing. $\Tilde{H}(\bar{k}_1,\, \bar{k}_2)$ denotes the hard
scattering kernel of the $S[ud]\overline{S[ud]} \,\to\, \bar{c}c$
subprocess. The  $p \,\to\, \overline{D^0}$ transition is written as
\begin{equation}
 \int \frac{d^4 z_1}{(2\pi)^4} \, e^{\imath \bar{k}_1 \cdot z_1} \,
 \langle \overline{D^0} :\, p^\prime \mid \mathcal{T} \Psi^c(-z_1/2) \Phi^{S[ud]}(+z_1/2) \mid p :\, p,\,\mu \rangle \,,
\label{eq:p-to-Dbar-transition-matrix-element}
\end{equation}
which is a Fourier transform of a hadronic matrix element of a
time-ordered, bilocal  product of a $c$-quark operator and an
$S[ud]$-diquark operator ($\mathcal{T}$ denotes the time-ordering of
the fields). In Eq.~(\ref{eq:p-to-Dbar-transition-matrix-element})
$\Phi^{S[ud]}(+z_1/2)$ takes out the $S[ud]$ diquark that enters the
hard subprocess from the proton state $\mid p :\, p,\,\mu \rangle$
at space-time point $+z_1/2$. $\Psi^c(-z_1/2)$ reinserts the
produced $\bar{c}$ quark into the remainders of the proton at
space-time point $-z_1/2$, which then gives the final $\mid
\overline{D^0} :\, p^\prime \rangle$ state. The  $\bar{p} \,\to\,
D^0$ transition is treated in an analogous way.

Given the heavy quark mass as a hard scale and taking into account
the  physically plausible assumption that the partons are almost on
mass-shell and their intrinsic transverse momenta are smaller than a
typical hadronic scale of the order of $1\,\text{GeV}$ the
transverse and minus (plus) components of the active (anti)parton
momenta are small as compared to their plus (minus) components. The
parton momenta are then approximatly proportional to the  hadron
momenta and one can perform the integrations over $\bar{k}_1^-$,
$\bar{k}_2^+$, $\mathbf{\bar{k}}_{1\perp}$ and
$\mathbf{\bar{k}}_{2\perp}$ in the convolution integral. Moreover,
the field operators in the hadronic matrix elements are forced to
have a light-like distance and thus the time ordering of the fields
in the hadronic matrix elements can be dropped. The $p \bar{p}
\,\to\, \overline{D^0} D^0$ amplitude then reads
\begin{equation}
\begin{split}
 M_{\mu\nu} \,=\, &
 \int d\bar{k}_1^+ \theta(\bar{k}_1^+) \int \frac{d z_1^-}{2\pi} e^{\imath \bar{k}_1^+ z_1^-}
 \int d\bar{k}_2^- \theta(\bar{k}_2^-) \int \frac{d z_2^+}{2\pi} e^{\imath \bar{k}_2^- z_2^+} \\
&\times \langle \overline{D^0} :\, p^\prime \mid \Psi^c(-z_1^-/2) \Phi^{S[ud]}(+z_1^-/2) \mid p :\, p,\,\mu \rangle
 \,\Tilde{H}(\bar{k}_1,\, \bar{k}_2)\, \\
&\times \langle D^0 :\, q^\prime \mid
\Phi^{S[ud]\,\dagger}(+z_2^+/2) \overline{\Psi}^c(-z_2^+/2) \mid
\bar{p} :\, q,\,\nu \rangle \,,
\label{eq:DHM-amplitude}
\end{split}
\end{equation}
with the $p \,\to\, \overline{D^0}$ transition matrix element
\begin{equation}
 \bar{p}^+ \, \int \frac{d z_1^-}{2\pi} e^{\imath \bar{k}_1^+ z_1^-}
 \langle \overline{D^0} :\, p^\prime \mid \Psi^c(-z_1^-/2) \Phi^{S[ud]}(+z_1^-/2) \mid p :\, p,\,\mu \rangle
\label{eq:LC-p-to-Dbar-transition-matrix-element}
\end{equation}
and an analogous one for the $\bar{p} \,\to\, D^0$ transition.
\section{Hadronic transition matrix elements}
\label{sec:Hadronic-transition-matrix-elements}

Using the same projection techniques as in
Ref.~\cite{Goritschnig:2009sq} we pick out the dominant components
of the $c$-quark field
\begin{equation}
 \Psi^c(-z_1^-/2) \,=\,
 \frac{1}{2 k_1^{\prime +}} \sum_{\lambda_1^\prime}
 v(k_1^\prime,\,\lambda_1^\prime) \left( \bar{v}(k_1^\prime,\,\lambda_1^\prime)
 \gamma^+ \Psi^c_\text{good}(-z_1^-/2) \right) \, .
\label{eq:c-quark-field-operator}
\end{equation}
Here $\Psi^c_\text{good}$ contains the dynamically independent
components of the $c$-quark field operator. With the help of
Eq.~(\ref{eq:c-quark-field-operator}) and an analogous one for the
antiquark the $p \bar{p} \,\to\, \overline{D^0} D^0$ amplitude
(\ref{eq:DHM-amplitude}) becomes
\begin{equation}
\begin{split}
 M_{\mu\nu} \,=\,
 & \frac{1}{4(\bar{p}^+)^2} \, \sum_{\lambda_1^\prime,\,\lambda_2^\prime} \,
 \int d\bar{x}_1 \, \int d\bar{x}_2 \, H_{\lambda_1^\prime,\,\lambda_2^\prime}(\bar{x}_1,\,\bar{x}_2) \,
 \frac{1}{\bar{x}_1-\xi} \, \frac{1}{\bar{x}_2-\xi} \, \\
 & \times \bar{v}(k_1^\prime,\,\lambda_1^\prime) \gamma^+ \bar{p}^+ \int \frac{d z_1^-}{2\pi} e^{\imath \bar{x}_1 \bar{p}^+ z_1^-}
 \langle \overline{D^0} :\, p^\prime \mid
 \Psi^c_\text{good}(-z_1^-/2) \Phi^{S[ud]}(+z_1^-/2)
 \mid p :\, p,\,\mu \rangle \\
 & \times \bar{q}^-  \int \frac{d z_2^+}{2\pi} e^{\imath \bar{x}_2 \bar{q}^- z_2^+}
 \langle D^0 :\, q^\prime \mid
 \Phi^{S[ud]\,\dagger}(+z_2^+/2) \overline{\Psi}^c_\text{good}(-z_2^+/2)
 \mid \bar{p} :\, q,\,\nu \rangle
 \gamma^- u(k_2^\prime,\,\lambda_2^\prime) \,,
\label{eq:DHM-amplitude-simplified}
\end{split}
\end{equation}
where we have defined
 $H_{\lambda_1^\prime,\,\lambda_2^\prime}(\bar{x}_1,\,\bar{x}_2) \,:=\,
 \bar{u}(k_2^\prime,\,\lambda_2^\prime) \,
 \Tilde{H}(\bar{x}_1\bar{p}^+,\,\bar{x}_2\bar{q}^-) \,
 v(k_1^\prime,\,\lambda_1^\prime)
$
and introduced the average momentum fractions $\bar{x}_1 =
\bar{k}_1^+/\bar{p}^+$ and $\bar{x}_2 = \bar{k}_2^-/\bar{q}^-$. In
order to make predictions one has to model the $p \,\to\,
\overline{D^0}$ and $\bar{p} \,\to\, D^0$ transition matrix
elements. We will do that by means of an overlap formalism in terms
of light-cone wave functions, which has been first developed in
Ref.~\cite{Diehl:2000xz} to represent the generalized parton
distributions of the proton.
We consider the proton and the $\overline{D^0}$ as $\mid S[ud] \, u
\, \rangle$ and $\mid \bar{c}\, u \,\rangle$ (bound)states,
respectively. In addition we assume the bound-state wave functions
to be pure s-wave, such that the parton helicities have to add up to
the total hadron helicity. Thus, the $u$ quark inside the proton has
to have the same helicity as the proton itself and the $\bar{c}$ 
and $u$ quark inside the $\overline{D^0}$ have to have opposite
helicities. The corresponding bound-state (light-cone) wave
functions of the proton and the $\overline{D^0}$ are  denoted by
$\psi_p$ and $\psi_D$, respectively. Those wave functions do not
depend on the total hadron momentum but only on the relative parton
momenta with respect to the parent hadron momentum. Working out the
wave-function overlaps as in Ref.~\cite{Goritschnig:2012vs} and
inserting the resulting $p \,\to\, \overline{D^0}$ and $\bar{p}
\,\to\, D^0$ transition matrix elements into
Eq.~(\ref{eq:DHM-amplitude-simplified})  we obtain
\begin{equation}
\begin{split}
 M_{\mu\nu} \,=\,
 & 2 {\mu\nu} \,
 \int d\bar{x}_1 \, \int d\bar{x}_2 \, H_{-\mu,\,-\nu}(\bar{x}_1,\,\bar{x}_2) \,
 \frac{1}{\sqrt{\bar{x}_1^2-\xi^2}} \, \frac{1}{\sqrt{\bar{x}_2^2-\xi^2}} \, \\
 & \times \int \frac{d^2\bar{k}_\perp}{16\pi^3}
    \psi_D(\hat{x}^\prime(\bar{x}_1,\,\xi),\,\mathbf{\hat{k}}^\prime_\perp(\mathbf{\bar{k}}_\perp,\,\bar{x}_1,\,\xi))
    \psi_p(\tilde{x}(\bar{x}_1,\,\xi),\,\mathbf{\tilde{k}}_\perp(\mathbf{\bar{k}}_\perp,\,\bar{x}_1,\,\xi)) \\
 & \times \int \frac{d^2\bar{l}_\perp}{16\pi^3}
    \psi_D(\hat{y}^\prime(\bar{x}_2,\,\xi),\,\mathbf{\hat{l}}^\prime_\perp(\mathbf{\bar{l}}_\perp,\,\bar{x}_2,\,\xi))
    \psi_p(\tilde{y}(\bar{x}_2,\,\xi),\,\mathbf{\tilde{l}}_\perp(\mathbf{\bar{l}}_\perp,\,\bar{x}_2,\,\xi)) \,.
\label{eq:DHM-amplitude-as-overlap}
\end{split}
\end{equation}
Quantities with a tilde (hat) relate to a frame where the incoming (outgoing) hadron
has vanishing $x$- and $y$-momentum component.\footnote{With Eqs.~(\ref{eq:DHM-amplitude-as-overlap}) and (\ref{eq:DHM-amplitude-peaking-approximation}) we correct Eqs.~(56) and (57)  in Ref.~\cite{Goritschnig:2009sq} by a missing factor 4. Correspondingly the cross sections given in Ref.~\cite{Goritschnig:2009sq} have to be multiplied with a factor 16.}

\section{\lq\lq Peaking approximation\rq\rq\ and hard scattering amplitude}
\label{sec:peaking-approximation-and-hard-scattering-amplitude}

The wave function for the heavy $D$-meson is  strongly peaked around
$x_0 \approx m_c/M$ with respect to its momentum-fraction dependence
\cite{Goritschnig:2012vs,Goritschnig:2009sq}. This behaviour is also
reflected in the overlap representation of the hadronic transition
matrix elements. It means that the kinematical regions close to the
peak position contribute most to the $\bar{x}_i$ integrations in
Eq.~(\ref{eq:DHM-amplitude-as-overlap}). Thus it is justified to
replace the momentum fractions appearing in the hard scattering
amplitude by the value of the peak position. After doing that it is
possible to pull the hard subprocess amplitude out of the
convolution integral, which is then rendered solely to an integral
over the hadronic transition matrix elements. After applying this
peaking approximation the $p \bar{p} \,\to\, \overline{D^0} D^0$
amplitude simplifies to
\begin{equation}
\begin{split}
 M_{\mu\nu} \,=\,
 & 2 {\mu\nu} \, H_{-\mu,\,-\nu}(x_0,\,x_0) \, \Big[
   \int d\bar{x} \,
   \frac{1}{\sqrt{\bar{x}^2-\xi^2}}
   \int \frac{d^2\bar{k}_\perp}{16\pi^3} \\
 & \times \psi_D(\hat{x}^\prime(\bar{x}_1,\,\xi),\,\mathbf{\hat{k}}^\prime_\perp(\mathbf{\bar{k}}_\perp,\,\bar{x}_1,\,\xi)) \,
    \psi_p(\tilde{x}(\bar{x}_1,\,\xi),\,\mathbf{\tilde{k}}_\perp(\mathbf{\bar{k}}_\perp,\,\bar{x}_1,\,\xi)) \Big]^2 \,.
\label{eq:DHM-amplitude-peaking-approximation}
\end{split}
\end{equation}

The hard $S[ud]\overline{S[ud]} \,\to\, \bar{c}c$ amplitudes can now
be calculated by applying the usual Feynman rules augmented with the
Feynman rules  for diquarks
\cite{Anselmino:1987vk,Anselmino:1987gu}; we obtain
\begin{equation}
\begin{split}
& H_{++} \,=\, + \, 4\pi\alpha_s(x_0^2 s) \, F_s(x_0^2 s) \, \frac{4}{9} \frac{2M}{\sqrt{s}} \, \cos\theta \,,\quad
  H_{+-} \,=\, - \, 4\pi\alpha_s(x_0^2 s) \, F_s(x_0^2 s) \, \frac{4}{9} \, \sin\theta \,, \\
& H_{--} \,=\, - \, 4\pi\alpha_s(x_0^2 s) \, F_s(x_0^2 s) \, \frac{4}{9} \frac{2M}{\sqrt{s}} \, \cos\theta \,,\quad
  H_{-+} \,=\, - \, 4\pi\alpha_s(x_0^2 s) \, F_s(x_0^2 s) \, \frac{4}{9} \, \sin\theta \,.
\label{eq:hard-subprocess-amplitudes}
\end{split}
\end{equation}
The form factor $F_s(x_0^2 s)$ accounts for the composite nature of
the diquark at  the $SgS$-vertex~\cite{Kroll:1993zx}.
\section{Modelling the $p \,\to\, \overline{D^0}$ transition}
\label{sec:modelling-the-hadronic-transitions}

In order to end up with an overlap representation of the $p \,\to\,
\overline{D^0}$ transition we have to specify the valence Fock state
light-cone wave functions for the proton and the $\overline{D^0}$.
According to Refs.~\cite{Goritschnig:2012vs,Kroll:1988cd} we take
\begin{equation}
 \psi_p(\tilde{x},\,\mathbf{\tilde{k}}_\perp) \,=\,
 N_p \, x \, e^{- a_p^2 \frac{\mathbf{\tilde{k}}_\perp^2}{\tilde{x}(1-\tilde{x})}}
\quad\text{and}\quad
 \psi_D(\hat{x}^\prime,\,\mathbf{\hat{k}}_\perp^\prime) \,=\,
 N_D \, e^{- a_D^2 \frac{\mathbf{\hat{k}}_\perp^{\prime 2}}{\hat{x}^\prime(1-\hat{x}^\prime)}} \,
  e^{- a_D^2 M^2 \frac{(\hat{x}^\prime-x_0)^2}{\hat{x}^\prime(1-\hat{x}^\prime)}}
\label{eq:LCWFs}
\end{equation}
as light-cone wave functions for the proton and the
$\overline{D^0}$, respectively. The light-cone wave function for the
$\overline{D^0}$ generates the peak around $x_0$ with the help of
the mass exponential. Each of the wave functions has two free model
parameters, the normalization constant $N_{p/D}$ and the transverse
size parameter $a_{p/D}$. For the proton we choose $a_p =
1.1\,\text{GeV}^{-1}$ and $N_p = 61.8\,\text{GeV}^{-2}$ which
amounts to $\langle \mathbf{k}_\perp^2 \rangle_{p}^{1/2} =
280\,\text{MeV}$ and the valence-Fock-state probability $P_p = 0.5$.
For the $\overline{D^0}$ we take $N_D = 55.2\,\text{GeV}^{-2}$ and
$a_p = 1.1\,\text{GeV}^{-1}$, which leads to $f_D = 206\,\text{MeV}$
(cf. Ref.~\cite{PDG}) and the valence-Fock-state probability $P_D =
0.9$.
In Fig.~\ref{fig:overlap} we show results for the overlap integral
occurring within the squar- function parametrization introduced above.
\section{Cross Sections}
\label{sec:cross-section}

\begin{figure*}
  \includegraphics[width=0.45\textwidth]{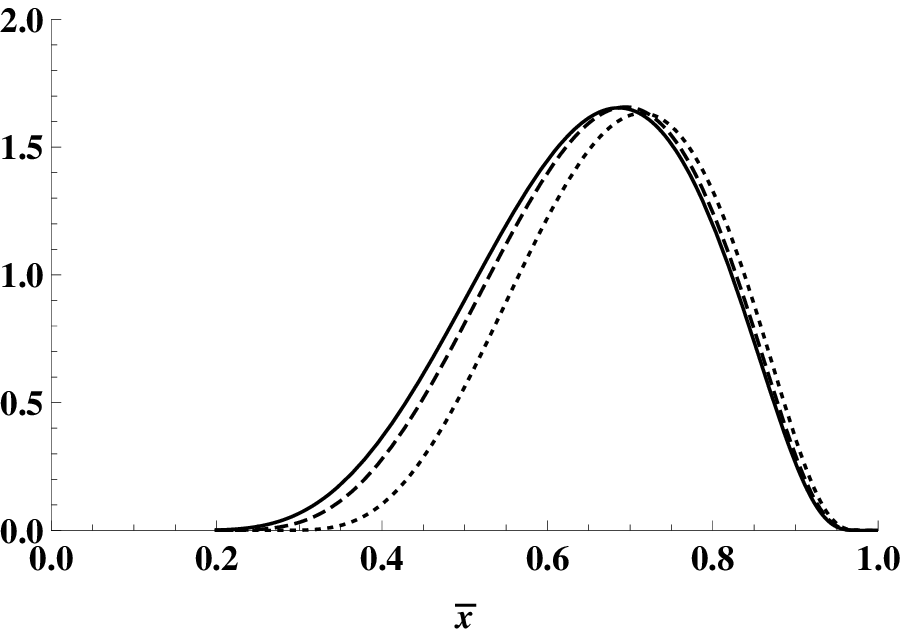}\hfill
  \includegraphics[width=0.45\textwidth]{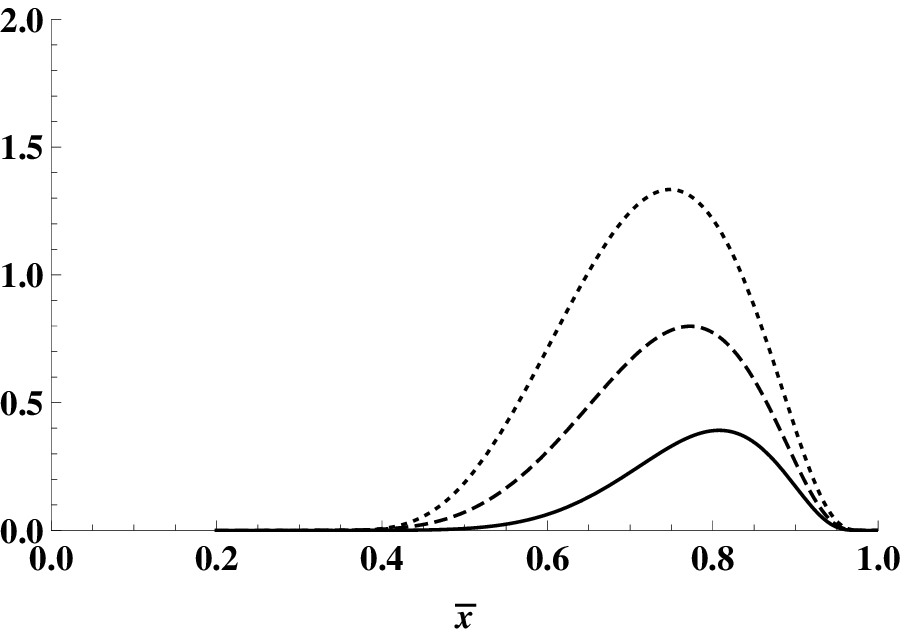}
\caption{The wave function overlap ocurring in Eq.~(\ref{eq:DHM-amplitude-peaking-approximation})
versus $\bar{x}$ at CMS scattering angles $\theta=0^\circ$ (left)
and $\theta=90^\circ$ (right) for $s=10, 20, 30$~GeV$^2$ (dotted,
dashed, solid).}\label{fig:overlap}
\end{figure*}

\begin{figure*}
 \includegraphics[width=0.45\textwidth]{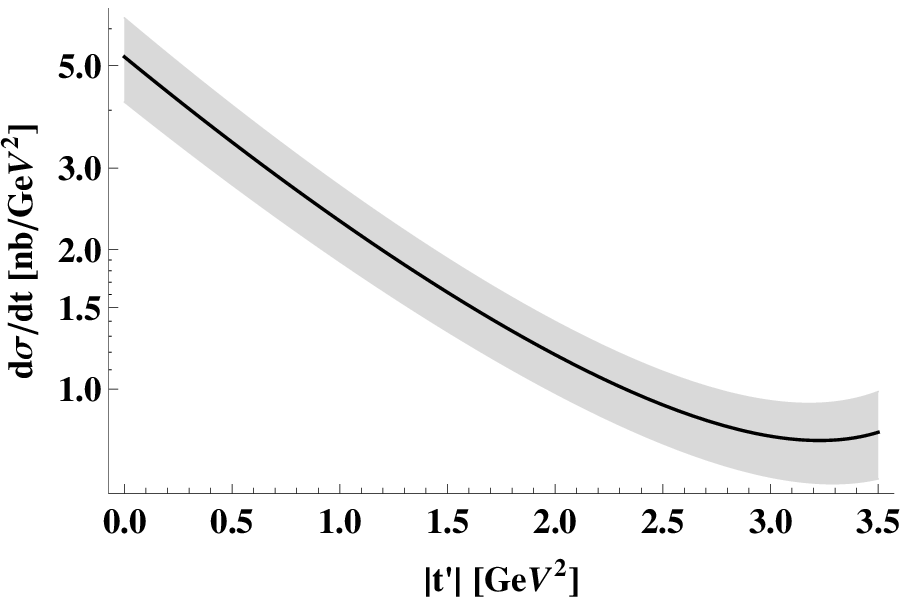}\hfill
 \includegraphics[width=0.45\textwidth]{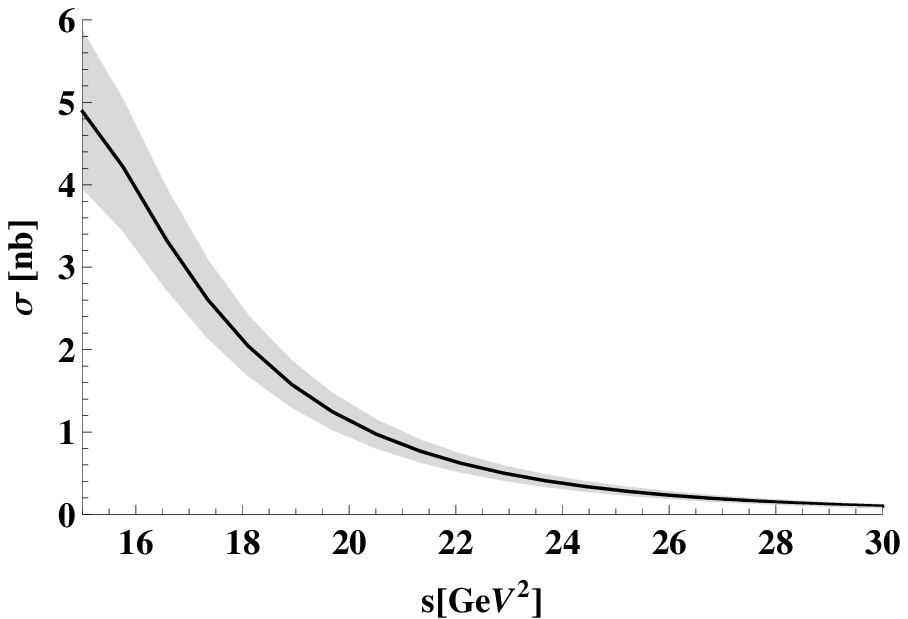}
\caption{The differential cross section
$d\sigma_{p\bar{p}\rightarrow\overline{D^0}D^0}/dt$ for
$s=15\,\text{GeV}^2$ (left) as a function of $|t^\prime|$ and the
integrated cross section (right) as a function of $s$.}
\label{fig:dsdt}
\end{figure*}

The differential $p \bar{p} \,\to\, \overline{D^0} D^0$ cross section reads
\begin{equation}
\frac{d\sigma_{p \bar{p} \,\to\, \overline{D^0} D^0}}{dt} \,=\,
\frac{1}{16\pi} \, \frac{1}{s^2} \, \frac{1}{1-4m^2/s} \, \sigma_0
\quad\text{with}\quad \sigma_0 \,:=\, \frac14 \, \sum_{\mu,\,\nu}
\mid M_{\mu\nu} \mid^2 \,. \label{eq:diff-cross-section}
\end{equation}
We show $d\sigma_{p \bar{p} \,\to\, \overline{D^0} D^0}/dt$ versus
$\mid t^\prime \mid=\mid t-t_0 \mid$ ($t_0=t(\theta=0)$) in the left
panel of Fig.~\ref{fig:dsdt} for Mandelstam $s=15\,\text{GeV}^2$.
The differential cross section is strongly decreasing with
increasing $\mid t^\prime \mid$. This behaviour comes from the
decrease of the model overlap with increasing CMS scattering angle
$\theta$, cf. Fig.~\ref{fig:overlap}. Its decrease becomes even more
pronounced for higher values of Mandelstam $s$. In the right panel
of Fig.~\ref{fig:dsdt}  we show the integrated cross section
$\sigma_{p \bar{p} \,\to\, \overline{D^0} D^0}$ versus Mandelstam
$s$, whose magnitude is in the range of a few $\text{nb}$.

\section{Summary}
\label{sec:Summary}

We have investigated the process $p\, \bar{p} \,\to\, \overline{D^0}
\, D^0$ within a   double handbag approach where the hard scale  is
given by the heavy $c$-quark mass. We have argued that under
physically plausible assumptions the $p\, \bar{p} \,\to\,
\overline{D^0} \, D^0$ amplitude factorizes into a hard subprocess
on the partonic level and transition distribution amplitudes. To
model the latter we have constructed an overlap representation in
terms of hadronic light-cone wave functions. Our predictions for the
differential and integrated $p\, \bar{p} \,\to\, \overline{D^0} \,
D^0$ cross section should now be confronted with future experimental
data from the $\overline{\text{P}}\text{ANDA}$ experiment at FAIR.




\end{document}